# How much are LLMs changing the language of academic papers after ChatGPT? A multi-database and full text analysis[1]

Kayvan Kousha

Statistical Cybermetrics and Research Evaluation Group, Business School, University of Wolverhampton, UK. https://orcid.org/0000-0003-4827-971X

Mike Thelwall

School of Information, Journalism and Communication, University of Sheffield, UK. https://orcid.org/0000-0001-6065-205X m.a.thelwall@sheffield.ac.uk

This study investigates how Large Language Models (LLMs) are influencing the language of academic papers by tracking 12 LLM-associated terms across six major scholarly databases (Scopus, Web of Science, PubMed, PubMed Central (PMC), Dimensions, and OpenAlex) from 2015 to 2024. Using over 2.4 million PMC open-access publications (2021–July 2025), we also analysed full texts to assess changes in the frequency and co-occurrence of these terms before and after ChatGPT's initial public release. Across databases, delve (+1,500%), underscore (+1,000%), and intricate (+700%) had the largest increases between 2022 and 2024. Growth in LLM-term usage was much higher in STEM fields than in social sciences and arts and humanities. In PMC full texts, the proportion of papers using underscore six or more times increased by over 10,000% from 2022 to 2025, followed by intricate (+5,400%) and meticulous (+2,800%). Nearly half of all 2024 PMC papers using any LLM term also included underscore, compared with only 3%–14% of papers before ChatGPT in 2022. Papers using one LLM term are now much more likely to include other terms. For example, in 2024, underscore strongly correlated with pivotal (0.449) and delve (0.311), compared with very weak associations in 2022 (0.032 and 0.018, respectively). These findings provide the first large-scale evidence based on full-text publications and multiple databases that some LLM-associated terms are now being used much more frequently and together in academic writing. However, the results do not provide direct causal evidence and cannot distinguish between LLM-generated text, LLM-edited text, or broader adoption of LLM-associated writing or publishing styles. The rapid uptake of LLMs to support scholarly publishing is a welcome development reducing the language barrier to academic publishing for non-English speakers.

**Keywords:** Large Language Models (LLMs); ChatGPT; academic writing; research communication; scholarly publishing; AI-assisted writing

## Introduction

Large Language Models (LLMs) like ChatGPT have the capability to assist academic writing (Khalifa & Albadawy, 2024) such as for editing and proofreading (Lechien et al., 2024), drafting abstracts (Gao et al., 2023; Hwang et al., 2024), generating literature reviews (Kacena et al., 2024; Margetts et al., 2024), performing statistical analyses (Huang et al., 2024), and even formulating research hypotheses (Park et al., 2024). Although AI tools have been used in publishing and peer review before (Kousha & Thelwall,

[1] This article is an extended version of the paper presented at the ISSI 2025 Conference. It incorporates a large-scale analysis of over 2.4 million full-text articles from PubMed Central (PMC) from 2021-July 2025, adding new results on the frequency, co-occurrence, and correlations analysis of 12 LLM-associated terms.

2024), the adoption of LLMs in research communication is still evolving. For example, most ChatGPT acknowledgements (80%) in academic papers relate to text editing and proofreading, rather than to more complex tasks such as coding or data analysis (Kousha, 2024).

Several recent surveys show growing use of LLMs in academic research, particularly for writing and editing tasks. An Elsevier survey (n=2,284) found that 31% of researchers had used generative AI for research activities, with 93% finding it helpful for writing and reviewing papers (Elsevier, 2024). A Nature survey (n=1,600) reported that 47% of scientists considered AI "very useful" for academic tasks, and 55% believed it saved time and resources (Van Noorden & Perkel, 2023). An Oxford University Press global survey in 2024 (n=2,345) found that 76% of researchers had used AI tools, 67% reported clear benefits, and 27% were optimistic about AI's potential for data analysis and content discovery. However, about half expressed concerns about AI's future impact on academic research (OUP, 2024). A global 2024 survey of publishing academics (n=1,711) found that the top uses of generative AI in research were translation (13.5%), proofreading (13%), drafting (12.5%), and literature review support (12.5%) (Mohammadi et al., 2026), indicating that generative AI is most often used for writing support rather than complex research practices. In medical research, 58% of surveyed urologists (n = 456) reported using ChatGPT for academic writing (Eppler et al., 2024), while 24% of authors in the medical sciences (n=229) used LLMs for rephrasing, proofreading, or translation (Salvagno et al., 2024). However, a global survey of clinical researchers (n=226) found that although about 88% were aware of LLMs, only 19% had used them in publications, mostly for grammar and formatting, and most did not acknowledge their use (Mishra et al., 2024).

Several studies have attempted to estimate the prevalence of LLM-associated language in academic publications. An early study in 2023 compared 9,953 GPT-3.5-generated abstracts with 10,000 human-written PubMed abstracts and identified 21 terms, including *delve*, *underscore*, and *intricate*, that were significantly more frequent in AI-generated texts, suggesting that LLM-generated language patterns might influence academic writing styles (Juzek & Ward, 2025). Gray (2024) found that just over 1% of 2023 papers (around 60,000) included LLM-associated terms such as *meticulously, innovatively, pivotal,* and *intricate*. Liang et al. (2024) reported that 17.5% of computer science preprints and 6.3% of Nature journal papers contained AI-modified content based on terms including *realm, intricate, showcasing,* and *pivotal*. In biomedical sciences, Kobak et al. (2024) found that the prevalence of LLM-associated terms such as *delves, showcasing,* and *underscores* in PubMed abstracts rose to 10% by 2024. Similarly, Uribe and Maldupa (2024) reported that LLM-associated terms (*delve, commendable, meticulous*, and *innovative*) in dental research indexed in PubMed increased from 47.1 to 224.2 papers per 10,000. Using AI-detection tools, Picazo-Sanchez and Ortiz-Martin (2024) estimated that 10% of 45,000 papers published between December 2022 and February 2023 were probably written with ChatGPT help. A study of 30,000 abstracts from 25 leading economics journals (2001–2024) also found a rise in LLM-associated terms, such as *delve, crucial*, and *intricate* (from 2.8% in 2023 to 6.7% in 2024), confirming the growing use of LLMs in this area (Feyzollahi & Rafizadeh, 2025). Another study analysed 2.4 million arXiv abstracts (2010–2024) using four readability metrics and found a steady decline in readability after the release of ChatGPT, suggesting that its growing use may be contributing to increased linguistic complexity in abstracts (Alsudais, 2025). Finally, an analysis of research article titles in Scopus (2015–2024) identified a sharp increase in AI-related terms such as *unveiling*, *enhancing*, and *exploring*, especially after 2023, indicating a growing influence of generative AI on the phrasing of article titles (Comas-Forgas, Koulouris, & Kouis, 2025).

Despite the above extensive findings, most previous studies have been limited to specific fields (e.g., computer science, dentistry, economics), early datasets (2023 to early 2024), small keyword lists, or only analysed abstracts. This study addresses these limitations by using much larger datasets from six major scholarly databases (Scopus, Web of Science [WoS], PubMed, PMC, Dimensions, and OpenAlex) and tracking 12 LLM-associated terms across a longer timeframe (2015–July 2025). Unlike earlier work, this is the first large-scale full-text analysis of over 2.4 million PMC publications, analysing not only the frequency of LLM-associated terms but also their repetition, co-occurrence patterns, and correlations within the main body of research papers. It also discusses disciplinary differences and provides additional analyses, including how the use of these terms compares with traditional academic terms and how often they appear in both retracted and published papers.

## Research questions

This study investigates how Large Language Models like ChatGPT may be influencing academic writing by tracking the usage patterns of 12 LLM-associated terms across scholarly databases. The first research question assesses how the prevalence and proportion of these terms have changed over time, particularly after the release of ChatGPT 3.5, and whether these trends differ between subjects. The second and third research questions investigate the frequency and co-occurrence of LLM-associated terms within the full text of academic publications to understand how frequently these terms were used and how they appear together in the main body of papers. The fourth research question analyses the correlations between LLM-associated terms in PMC full texts, providing new evidence on whether the associations between using these terms have increased after ChatGPT's release.

- **RQ1:** How has the use of LLM-associated terms in academic publications changed after ChatGPT public release and how does this vary across scholarly databases and broad subjects?
- **RQ2:** How has the frequency of LLM-associated terms changed in PMC full-text articles before and after the release of ChatGPT?
- **RQ3:** How have the co-occurrences between LLM-associated terms changed in PMC full-text articles before and after the release of ChatGPT?
- **RQ4:** To what extent are LLM-associated terms correlated with each other in PMC full texts before and after the release of ChatGPT?

# Methods

### Identifying LLM-associated terms

In this study, we investigated language change in academic writing before and after ChatGPT's public release (using GPT 3.5) in November 2022, using a combination of bibliometric and full-text databases. To do this, we tracked terms commonly associated with LLM-associated texts. These terms were either drawn from previous studies (Table 1) or identified through our own initial analysis.

For the latter, we extended the list of LLM-associated terms using an exploratory analysis of Scopus titles and abstracts in Environmental Science/Environmental Studies, which is large and had relatively high use of several previously reported LLM-associated terms. For example, for the query TITLE-ABS-KEY(underscore OR underscores OR underscored OR underscoring) limited to 2024–2025, Environmental Science contained 335,026 documents in total, and 12,200 matched the underscore variants (3.6%), which was higher than most other Scopus subject areas in the same period.

We started with a seed list of terms drawn from prior studies (Table 1) (e.g., underscore, delve, showcase, unveil, intricate, meticulous, pivotal) and used Scopus term-comparison outputs to identify additional words that were more common in Environmental Science titles/abstracts when the seed terms were present than when they were absent. This produced a long list of about 1,000 candidate associated terms. We then screened this list in two steps. First, we retained candidates that co-occurred with the seed-term set more often than expected using a $\chi^2$ test ($p < 0.01$). Second, we checked each remaining candidate using Scopus annual trend graphs and kept only those showing a clear step-change in frequency in 2024. This produced five additional terms (heighten, nuance, bolster, foster, interplay), which were added to the seven terms drawn from the literature, giving the final set of 12 terms used in all analyses. We note that some added terms (e.g., foster, heighten) can also be used in discipline-specific ways and the selection of LLM-associated terms is not comprehensive and is discussed further in the Limitations section.

Table 1. lists the final terms selected for analysis and sources from which they were identified. Although there is no direct causal evidence that these terms originate from LLMs, it is reasonable to hypothesise that recent increases in their usage may be influenced by LLM-associated text. This assumption is based on previous studies and the fact that these terms are generic in nature, lacking any clear alternative origin (unlike domain-specific terms such as "Covid-19" or explicit labels like "LLM").

Table 1: Selected LLM-associated terms and supporting studies

| LLM-associated term | Supporting studies |
|---|---|
| underscore[s/d/ing] | Kobak et al., 2024; Uribe & Maldupa, 2024; Juzek & Ward, 2025 |
| delve[s/d/ing] | Kobak et al., 2024; Uribe & Maldupa, 2024; Juzek & Ward, 2025 |
| showcase[s/d/ing] | Kobak et al., 2024; Liang et al., 2024; Uribe & Maldupa, 2024 |
| unveil[s/ed/ing] | Uribe & Maldupa, 2024; Comas-Forgas et al., 2025 |
| intricate[s/d/ing] | Gray, 2024; Liang et al., 2024; Uribe & Maldupa, 2024; Juzek & Ward, 2025 |
| meticulous[ly] | Gray, 2024; Uribe & Maldupa, 2024 |

| pivotal | Gray, 2024; Liang et al., 2024 |
|---|---|
| heighten[s/ed/ing] | Authors’ analysis from Scopus articles in the Environmental Studies |
| nuance[s/d] | Authors’ analysis from Scopus articles in the Environmental Studies |
| bolster[s/ed/ing] | Authors’ analysis from Scopus articles in the Environmental Studies |
| foster[s/d/ing] | Authors’ analysis from Scopus articles in the Environmental Studies |
| interplay[s/ed/ing] | Authors’ analysis from Scopus articles in the Environmental Studies |

**Database searches for term prevalence (2015–2024)**

To address the first research question, the selected LLM-associated terms were searched for separately in the titles, abstracts, and keywords of bibliometric databases (Scopus, WoS, and PubMed) and through unrestricted searches in three platforms that index both metadata and some full text (OpenAlex, Dimensions, and PMC). To capture variations of each term, OR operators were used in all databases (e.g., delve OR delves OR delved OR delving). The results were limited to articles, reviews, and proceedings papers published between 2015 and 2024 to analyse term usage over a decade in academic papers and to guard against changes since 2022 being part of a longer-term trend unrelated to LLMs. To account for the annual growth in publication volume (e.g., fewer papers in 2015 than in 2024), the results were normalised by dividing term counts by the total number of publications indexed per year in each database. This proportional approach improves on previous studies that only reported raw frequency counts. All searches were conducted on 20 December 2024 to minimise variations caused by daily additions to the databases.

**PMC full-text term analysis (2021–2025)**

The method used to answer RQ1 only captured the prevalence of LLM-associated terms across six databases based on a single occurrence per paper in titles, abstracts, keywords, or metadata. However, this approach does not assess how frequently these terms appear. To address RQ2 and RQ3, we hypothesised that the release of LLMs like ChatGPT not only increased the overall use of LLM-associated terms but also led to their more frequent use and co-occurrence within full texts after 2022.

To test this, we downloaded the full text of PMC open-access publications in XML format from 2021 to 23 July 2025, providing a collection of over 2.4 million publications. The PMC full-text analysis was conducted as a follow-up to the initial multi-database searches and was downloaded later, which is why this dataset extends into 2025. A program was written and added to the free Webometric Analyst software (https://github.com/MikeThelwall/Webometric_Analyst: Citations menu > PMC full text > Count frequency of text of string in all XML articles) to count the frequency of each term in the main body of each paper (excluding titles, abstracts, references, and supplementary sections). This approach allowed us to conduct a more detailed analysis of (a) the frequency of LLM-associated terms within full texts, (b) their co-occurrence patterns, and (c) correlations between terms before and after ChatGPT public launch.

Geometric means were used to calculate average usage of LLM-associated terms in the PMC full-text analysis (Figure 8). Geometric means were used instead of the arithmetic mean because it is a better central tendency indicator for highly skewed data, with many documents containing zero occurrences. Using the Webometric Analyst software, we fed predefined variants of each of the 12 LLM-associated terms (e.g., delve, delved, delves, delving). The software then counted how often these terms appeared in the main body of PMC full texts, excluding titles, abstracts, references, and supplementary sections. Using lower-case term variants also reduced the likelihood of counting proper names (e.g., Foster as a surname).

# Results

## Trends in LLM-associated terms across scholarly databases

*Proportion of publications with LLM-associated terms*

Figure 1 shows the proportion of academic publications containing six selected LLM-associated terms across six databases from 2015 to 2024; the remaining six terms are presented in Appendix A to improve readability. The results indicate a clear increase in the proportion of academic publications using these LLM-associated terms after the ChatGPT 3.5 public release in late November 2022.

In 2024, *underscore[s/d/ing]* was the most frequent term, occurring in about 20% of PMC open-access publications, rising from just 3.6% in 2022, which represents almost a 5.5× increase in only two years. The next most frequent term was *pivotal* at about 15% in PMC, and a similar pattern was observed in Dimensions publications with approximately 11% for *underscore* and 8% for *pivotal*. For Dimensions, *delve* also shows a significant increase from 0.46% in 2022 to 4.3% in 2024, almost a 9× increase. Terms such as *intricate, meticulous, foster*, and *interplay* also increased steadily between 2022 and 2024.

There were small increases in the percentage of documents containing any of the 12 terms across all databases in 2023 and much larger increases in nearly all cases in 2024. OpenAlex is a slight anomaly, with increases in 2023 but not 2024. This may reflect OpenAlex using the first date it finds for a publication (often a preprint) rather than the formal publication date, so it can appear about a year ahead of the others. Across databases, title/abstract/keyword searches for WoS and Scopus give similar patterns and, unsurprisingly, the highest percentages occur in sources that include some full text (PMC and Dimensions). This supports the idea that LLMs are not only used to produce or polish abstracts but may also be used for generating or editing the main text of papers, which we investigate in later sections.

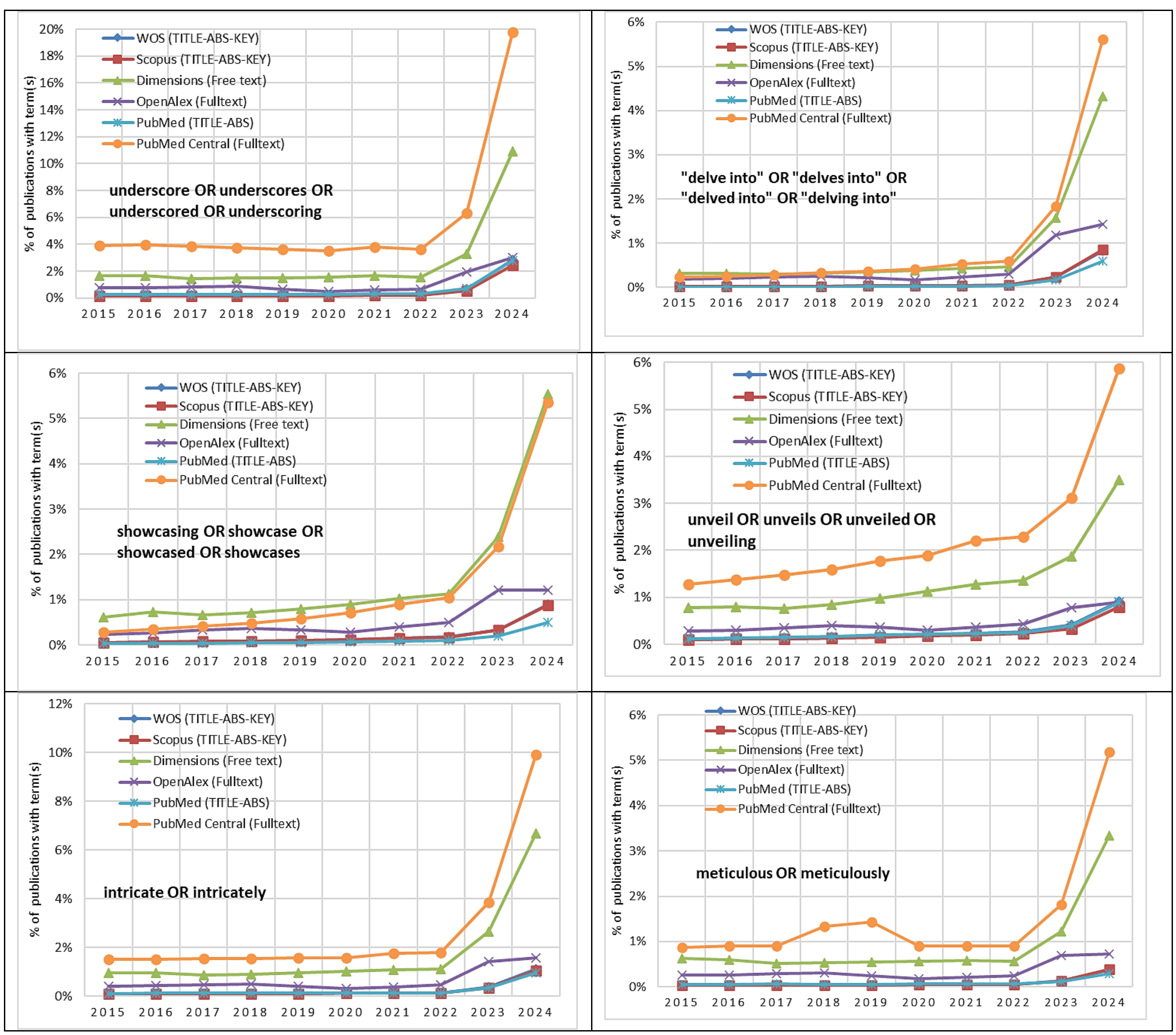

Figure 1. Percentage of academic publications containing six selected LLM-associated terms in six databases. The remaining six terms are shown in Appendix A to make the figures easier to read. Percentages are calculated separately for each year and are normalised by the total number of publications indexed in each database for that year.

*Growth in LLM-associated terms in academic publications (2022-2024)*

Figure 2 shows the relative growth of LLM-associated terms between 2022 and 2024 across databases. The terms *delve* and *underscore* had the highest increases: *delve* increased by up to 1,582% in Scopus and 1,331% in WoS, while *underscore* increased by 1,046% and 895%, respectively. Substantial growth was also observed in Dimensions and PMC for these terms, suggesting that these language changes are widespread across both abstract and full-text databases. The terms *intricate* and *meticulous* also showed strong growth between 2022 and 2024, increasing by 644% and 519% in WoS and by 736% and 613% in Scopus, respectively. In contrast, *interplay* and *foster* showed relatively modest increases, typically below 200% across most databases. Similarly, *nuanced* and *unveil* showed moderate growth, averaging around 150% to 250% depending on the source. This great variability in increases may reflect a range of factors, such as their initial rarity, whether they are similar to more academic terms that they have replaced, and how often they occur in non-academic texts (where LLMs presumably learn how to use them).

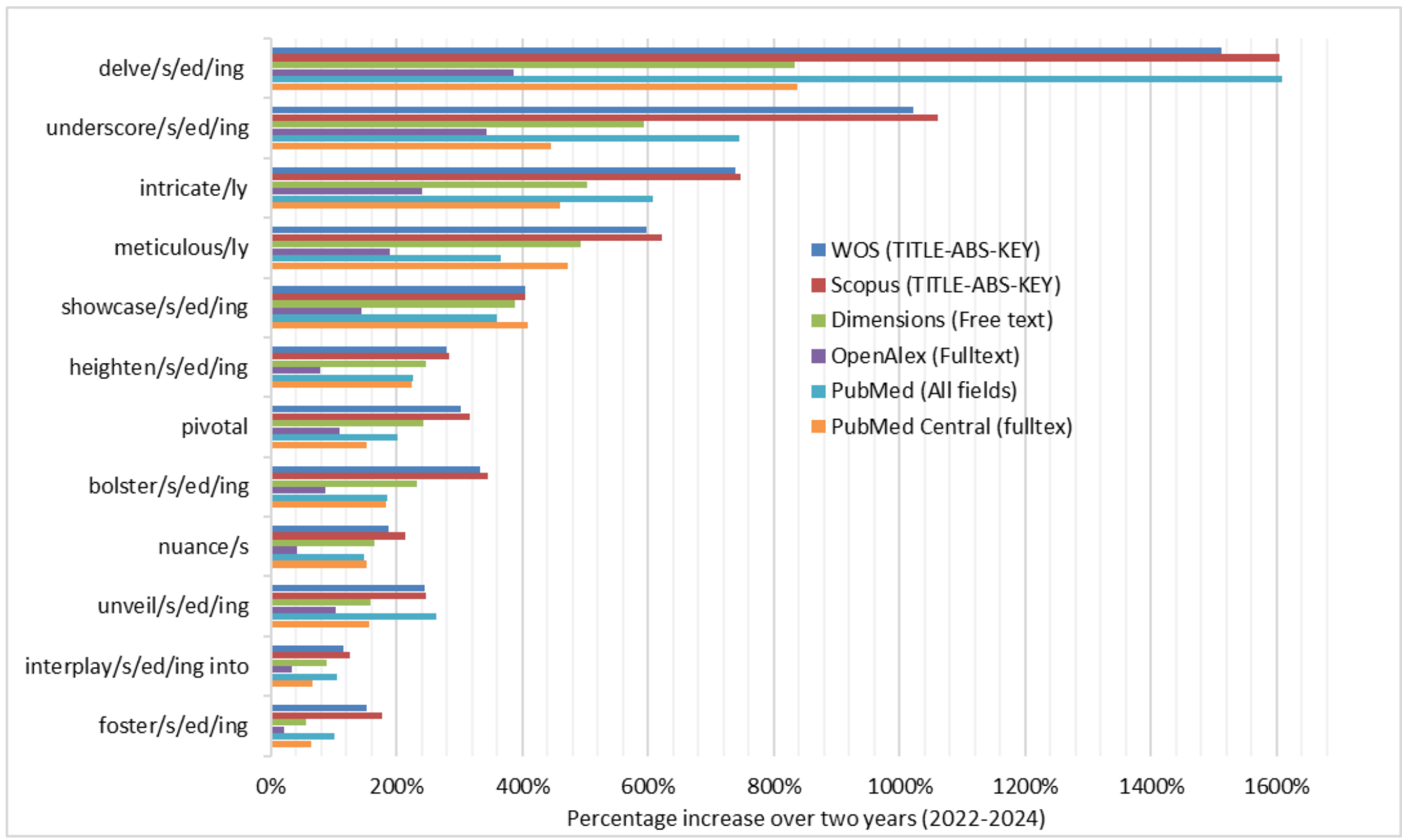

Figure 2. Percentage increases from 2022 to 2024 in the proportions of academic publications containing LLM-associated terms.

*Disciplinary differences in using LLM terms*

Figure 3 shows that the relative growth of LLM-associated terms between 2022 and 2024 varied substantially between Scopus subject areas. Among the fastest-growing terms, *delve* had the largest increases in Materials Science (+3,700%) and Chemical Engineering and Physics & Astronomy (both around +3,500%). *Underscore* had the highest growth in Computer Science (+3,580%), Engineering (+3,470%), and Mathematics (+3,300%), suggesting that these fields are adopting LLM-influenced language rapidly. *Meticulous* also had notable increases, rising most in Multidisciplinary research (+2,290%) and Decision Sciences (+1,750%). By contrast, some subject areas had slower uptake of LLM-associated terms, below +500% in fields such as Arts & Humanities, Social Sciences, and Psychology. These findings indicate that the use of LLM in academic writing is strongest in STEM fields, while growth has been much lower in the social sciences and arts and humanities. Note that the high percentage increases are possible because of the low values in 2022.

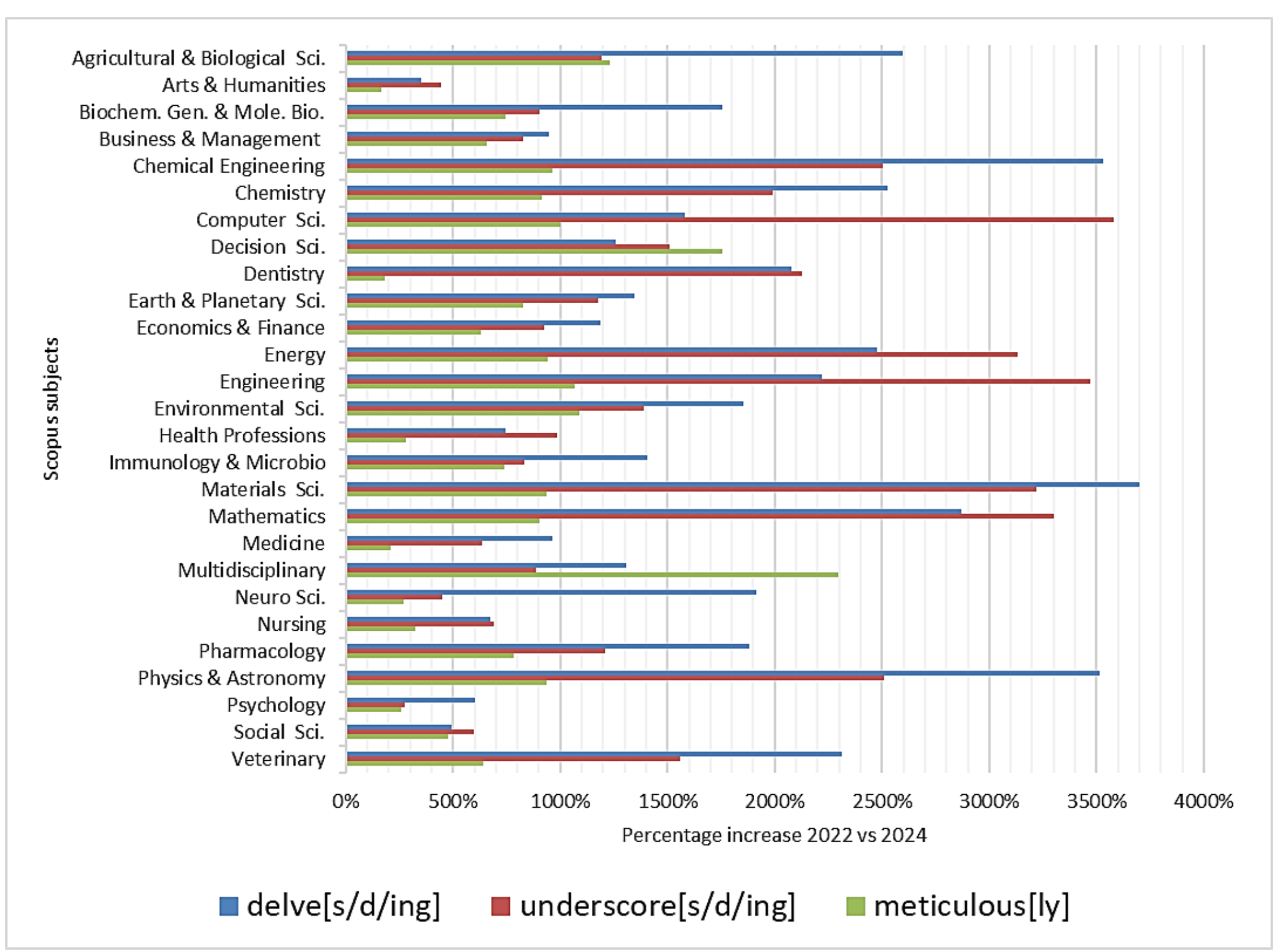

Figure 3. Percentage increase in the proportions of publications including three LLM-associated terms (*delve*, *underscore*, and *meticulous*) across 27 Scopus subject areas between 2022 and 2024.

The scatter plots in Figures 4 and 5 show a positive correlation between the percentages of publications using *delve[s/d/ing]* and *underscore[s/d/ing]* in 2022 and 2024 across most Scopus subject areas. This means that subjects where these terms were already used more in 2022 also tend to have the highest percentages in 2024 publications. However, there are some differences between subjects.

For *delve[s/d/ing]* (Figure 4), the highest 2024 percentages are in Arts & Humanities (1.67%), Social Sciences (1.46%), and Business, Management & Accounting (1.67%). These fields already used the term *delve* more before 2022, so they remain near the top in 2024. In contrast, Mathematics (0.72%), Dentistry (0.25%), and Physics & Astronomy (0.62%) have much lower percentages in 2024, even though their relative growth since 2022 is substantial because they started from very low levels. For example, use of *delve* in Mathematics publications rose from 0.02% in 2022 to 0.72% in 2024, an increase of more than 30 times.

A similar trend is observed for *underscore[s/d/ing]* (Figure 5). The highest 2024 percentages are in Environmental Science (3.84%), followed by Economics, Psychology, Social Sciences, and Business, Management & Accounting (all around 3.5%). In contrast, Mathematics (1.26%), Physics & Astronomy (1.49%), and Dentistry (1.83%) remained lower overall but still have significant increases compared with 2022 (e.g., Environmental Science rose from 0.26% in 2022 to 3.84% in 2024, a 15-fold rise).

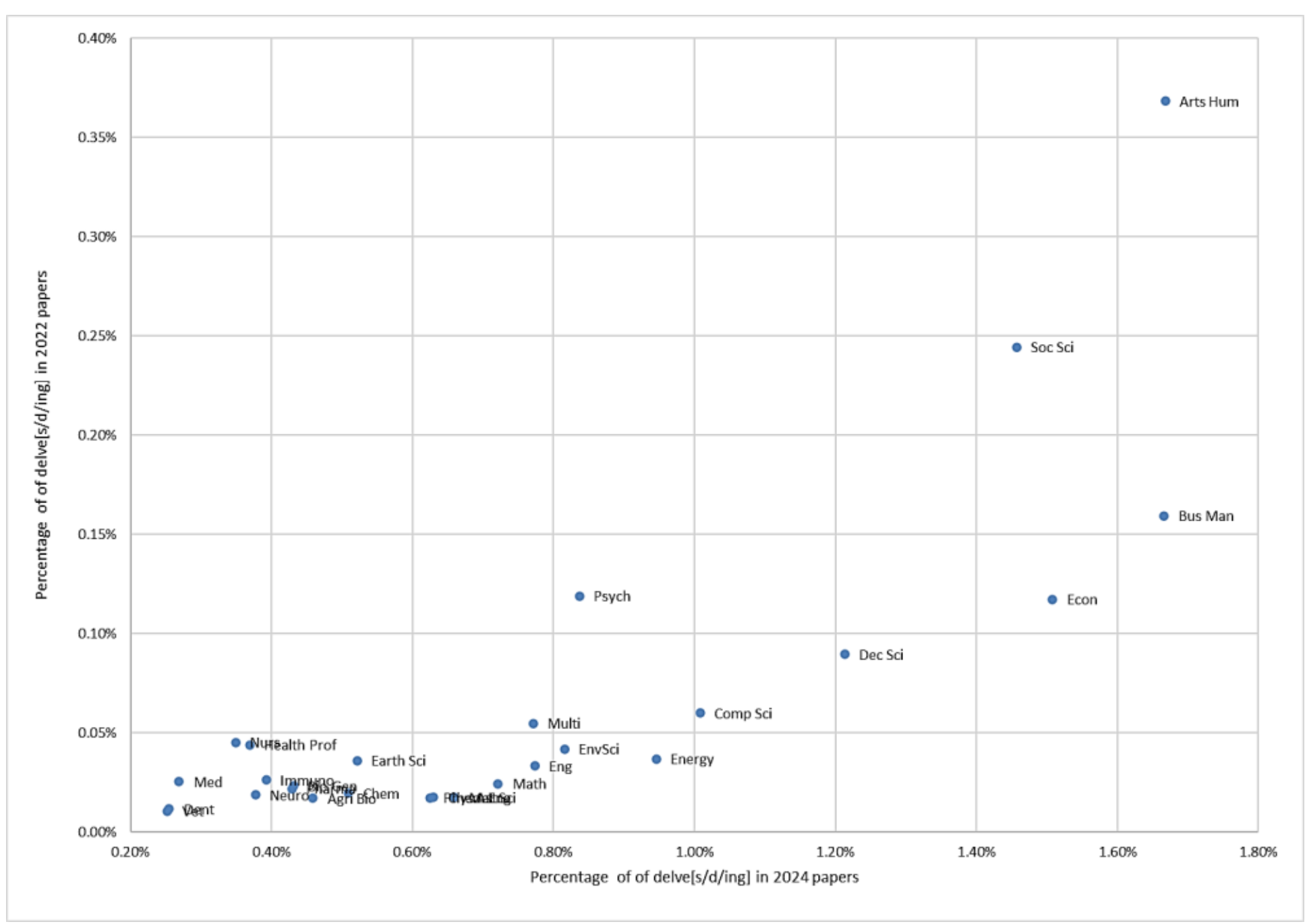

Figure 4. Scatter plot of the percentage of Scopus papers containing *delve[s/d/ing]* (2022 vs 2024) across 27 subjects.

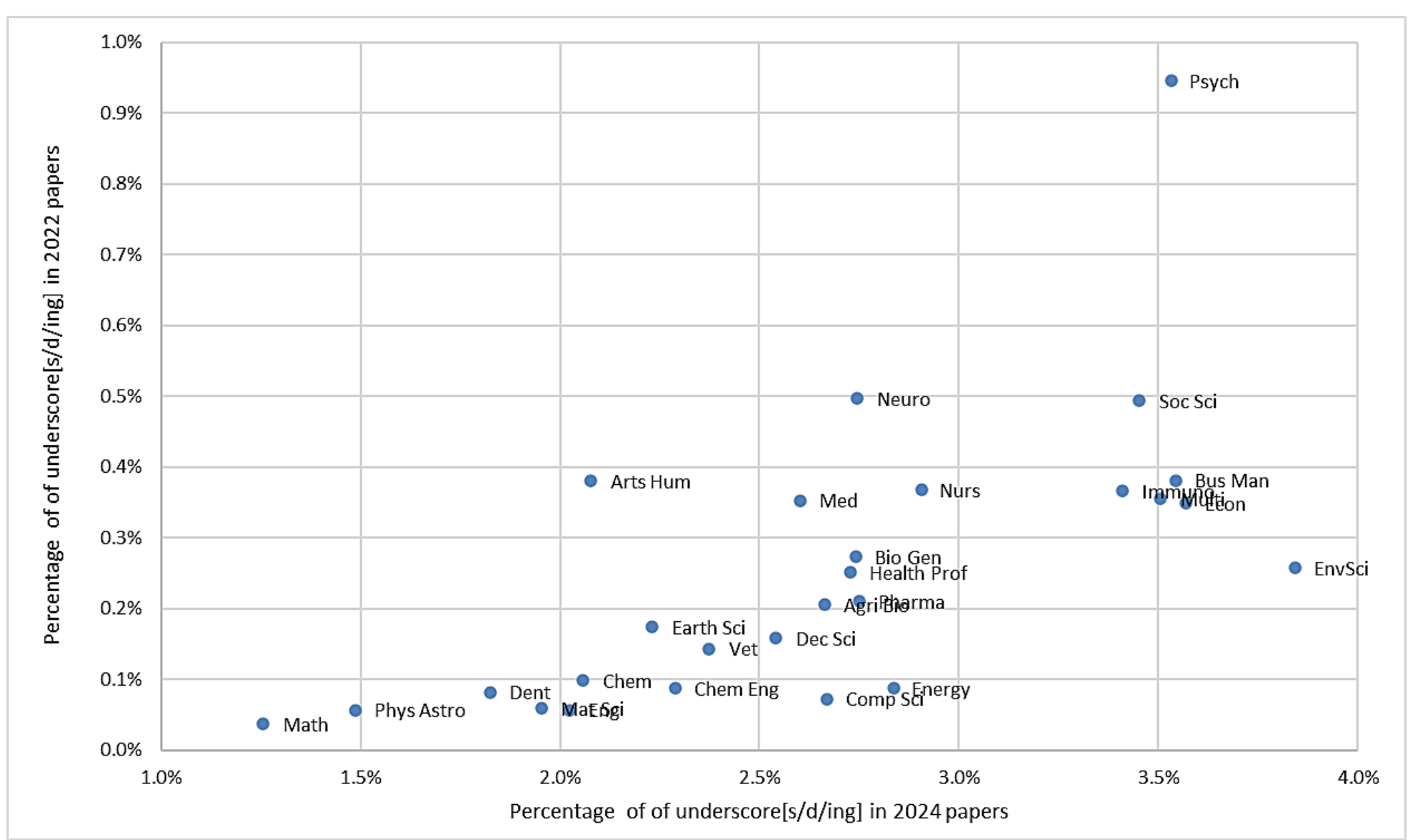

Figure 5. Scatter plot of the percentage of Scopus papers containing *underscore[s/d/ing]* (2022 vs 2024) across 27 subjects.

## PMC full-text analysis of LLM-associated terms

*Proportion of PMC full texts with LLM terms (2021–2025)*

Figure 6 shows the proportion (non-zero) of PMC publications mentioning at least one of the 12 LLM-associated terms in their full text between 2021 and 2025. The results indicate a sharp increase in the use of these terms after 2022, confirming the strong influence of LLM-associated language in biomedical publishing. *Underscore* shows the largest growth in full texts from about 3% in 2022 to nearly 30% in 2025 (10-fold rise). *Interplay* and *pivotal* also had substantial increases from 4% and 5% in 2022 to 11% and 15% in 2025, respectively. Other terms also had obvious growth after ChatGPT's release in 2022. The error bars for all terms do not overlap when comparing the periods before (2021–2022) and after the ChatGPT public release (2023–2025), so the observed increases are statistically significant at the 95% confidence level.

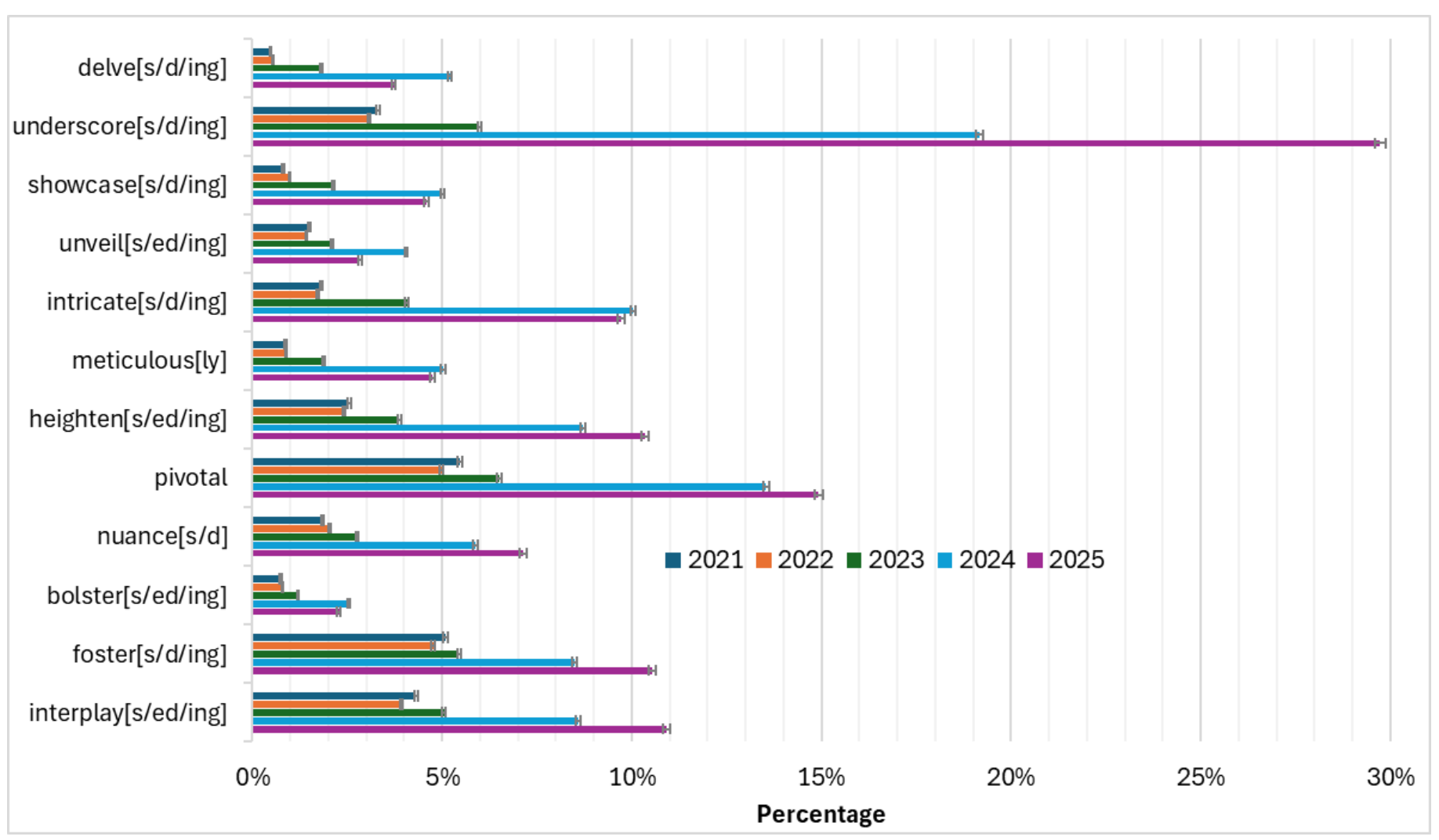

Figure 6. Proportion of PMC full-text articles mentioning each LLM-associated term at least once (2021–2025) with 95% confidence intervals.

However, Figure 7 shows significant differences among the main document types in PMC for the use of "underscore" between 2022 and 2025. In 2022, only 4.2% of review papers, 3.2% of research papers, and 1.1% of case reports used the term. By 2025, these proportions increased significantly to 44% for reviews, 29.9% for research papers, and 27.9% for case reports. This means that the use of "underscore" became 10 times higher in reviews, 9 times higher in research papers, and about 24 times higher in case reports (27.9 ÷ 1.1 = 24.2). One explanation for the higher proportion of review papers using "underscore" is that review articles tend to be much longer and cover broader topics, which naturally increases the chances of using LLM-associated terms more within the same paper and may partly reflect document length rather than term choice.

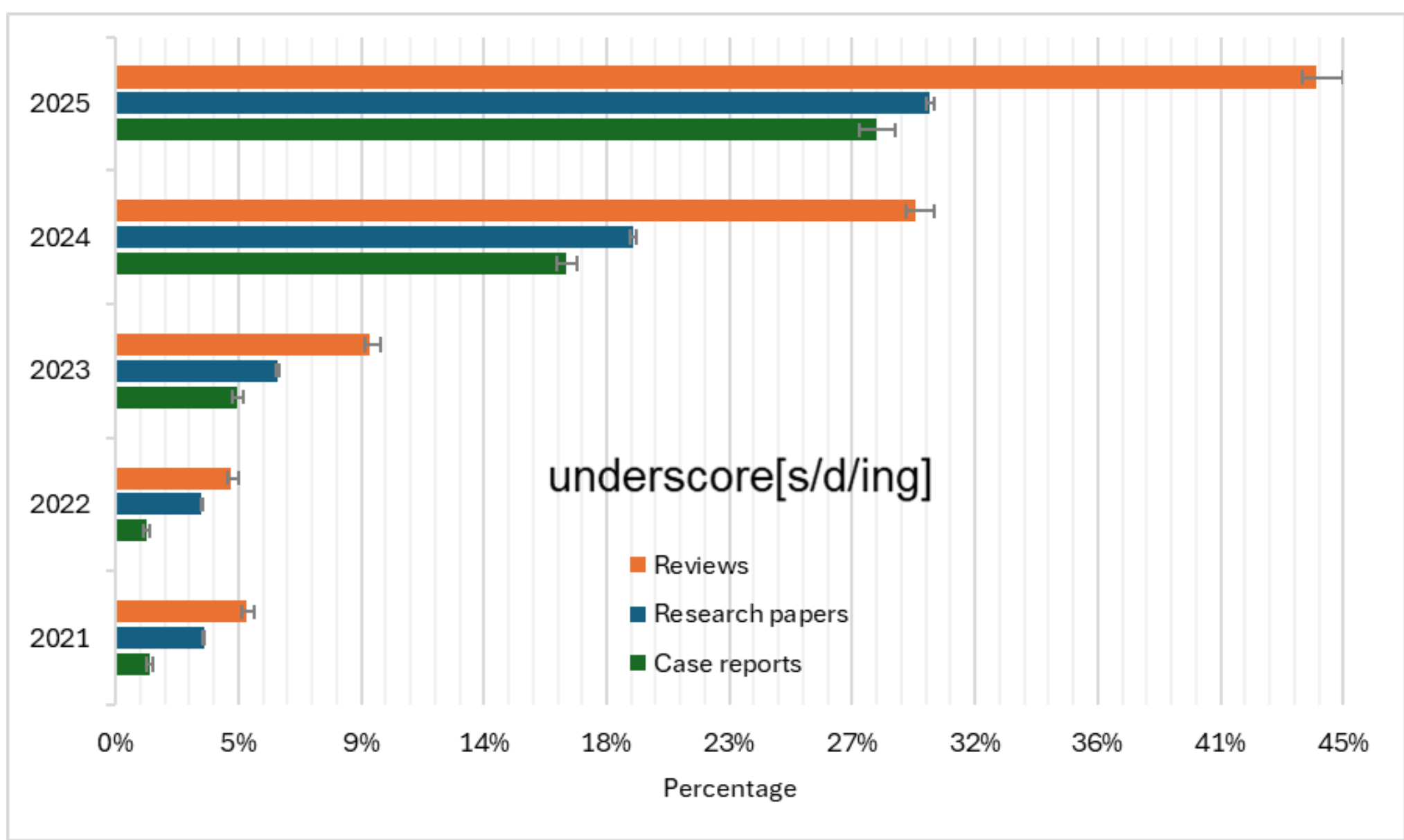

Figure 7. Percentage of PMC full-text publications mentioning *underscore[s/d/ing]* at least once (2021–2025) by main document type (research papers, review articles, and case reports) with 95% confidence intervals.

*Average usage of LLM-associated terms in PMC full texts (2021–2025)*

Figure 8 shows the geometric mean for each LLM-associated term across PMC full-text articles that mention the term at least once, between 2021 and 2025. This metric reflects the average use of each term within papers that already include it. The results indicate that *underscore* had the largest increase in average usage, rising from a geometric mean of 1.22 in 2022 to 1.76 in 2025. This suggests that not only are more papers mentioning *underscore* (as shown in Figure 6), but those that use it tend to use the term more often after 2022.

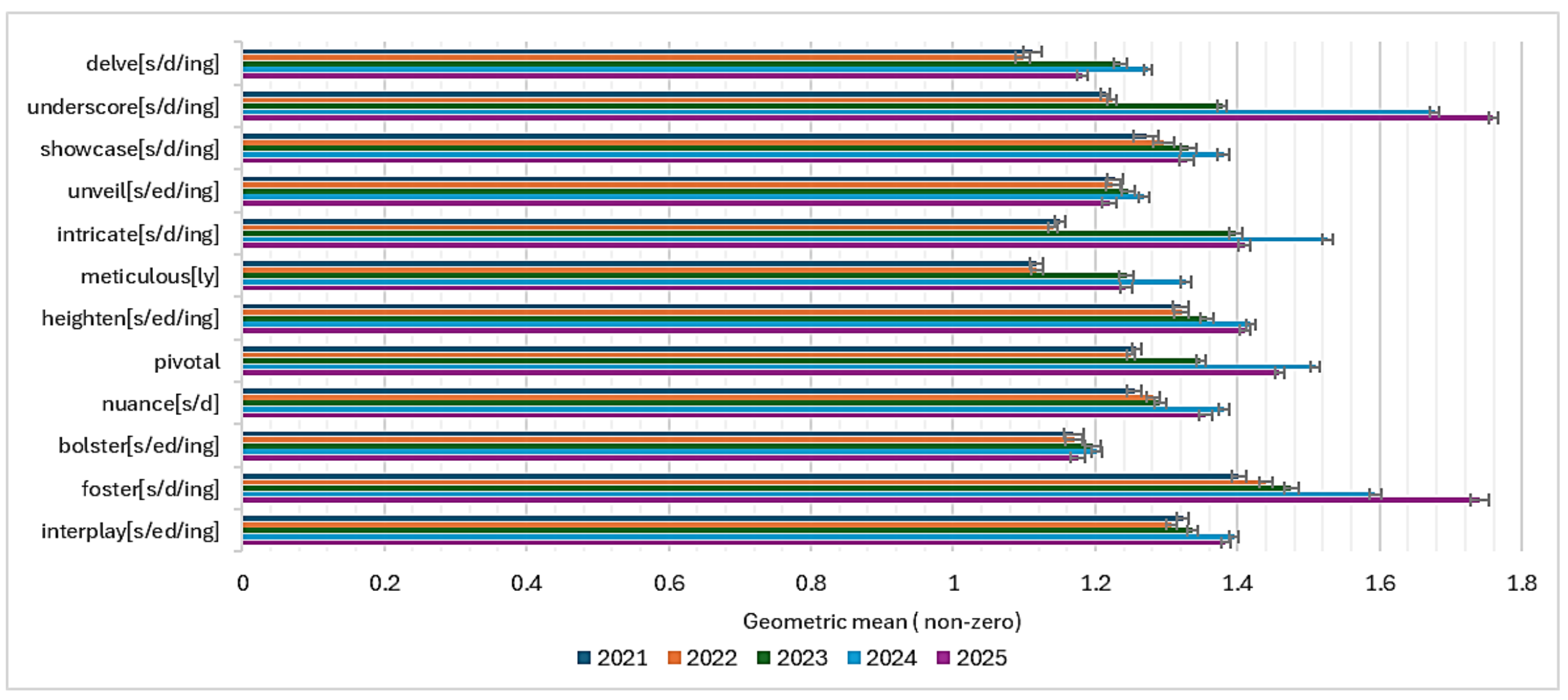

Figure 8. Geometric mean number of LLM-associated terms per PMC full-text article mentioning each term at least once (2021–2025) with 95% confidence intervals.

*Frequency of LLM term in PMC full texts (2022 vs 2025)*

Adding more detail to the results above, there were sharp increases in the frequent use of individual LLM-associated terms within PMC full-text publications between 2022 and 2025 (Figure 9), indicating that frequent use has become more common. The most substantial increase is for *underscore*, where the proportion of papers using the term six or more times rose from 0.013% in 2022 to 1.37% in 2025, an increase of over 10,000% (calculated as ((1.372 − 0.013) / 0.013) × 100). Similarly, *intricate* and *meticulous* increased by over 5,400% and 2,800%, respectively, for papers where the terms appeared six or more times. However, *delve* had its highest increase in usage up to five times per article between 2022 and 2025, suggesting that it is less likely to be used six or more times compared with terms such as *underscore* or *intricate*. This suggests that delve is used more selectively in academic writing rather than being repeatedly used within the full texts, although it appears more frequently in abstracts compared with other terms, as shown in Figure 2.

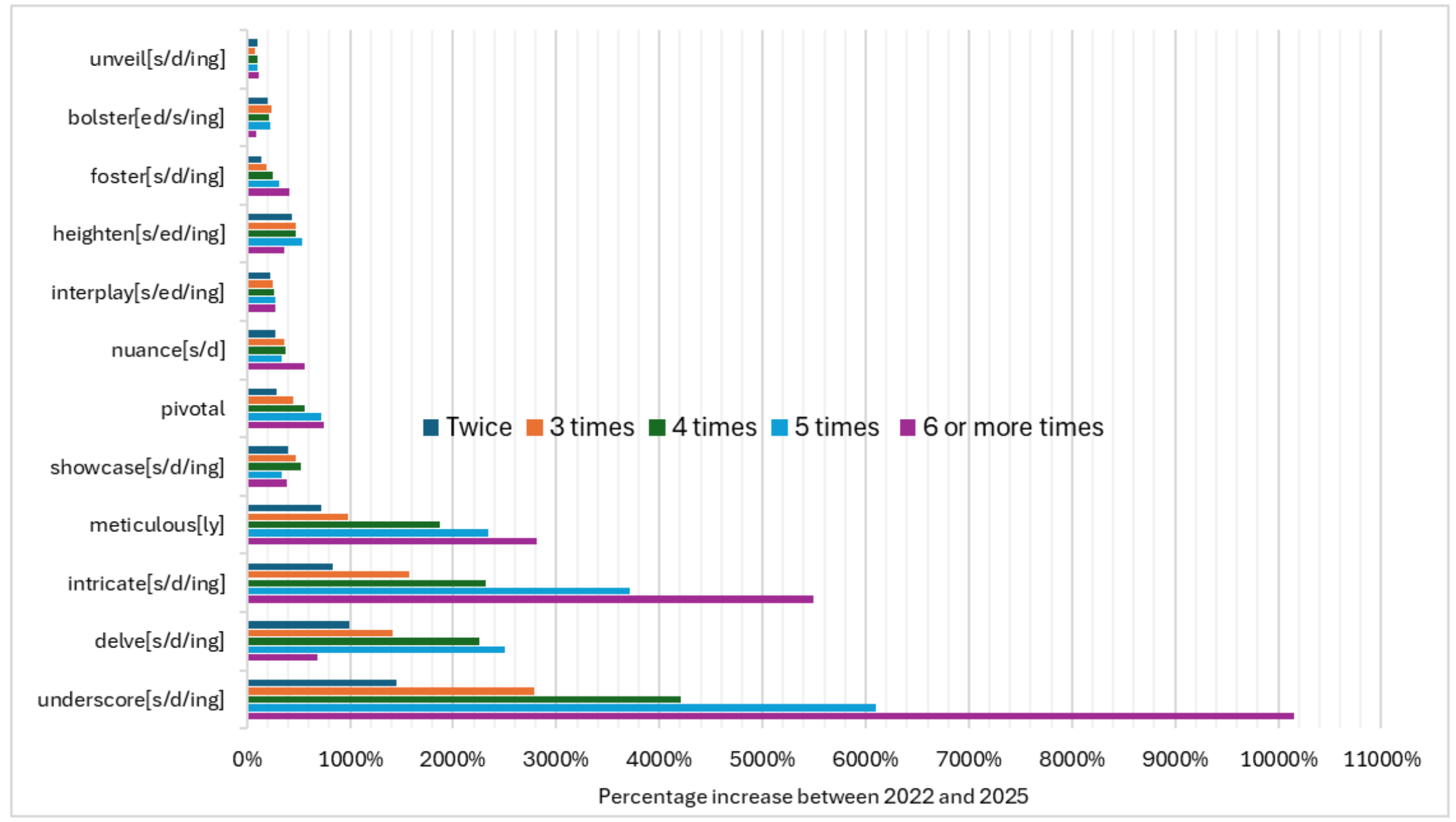

Figure 9. Percentage increase in the repeated use of LLM-associated terms within PMC full-text publications between 2022 and 2025 based on their occurrences (2 to 6+ times) per article.

*Co-occurrence of LLM-associated terms in PMC full texts (2022 vs 2024)*

To assess how often LLM-associated terms are used together, we calculated the conditional probability of two terms appearing within the same PMC full-text publication. For example, Figure 11 shows that in 2024, 59.3% of PMC papers mentioning *delve* (37,534 in total) also mentioned *underscore* (22,241 papers). However, only 16.1% of papers mentioning *underscore* (138,183 in total) also mentioned *delve* (22,241 papers). This shows that while papers using *delve* often also include *underscore*, the reverse is less common because *underscore* is used much more widely within PMC papers, and only about 16% of them mention *delve* as well. Therefore, the results differ within rows or columns because the probabilities are calculated relative to total numbers of papers for each term.

Figures 10 and 11 show how these co-occurrences patterns changed before and after the release of ChatGPT. In 2024, co-occurrence probabilities increased substantially compared with 2022, showing that papers using one LLM-associated term were significantly more likely to include others in the same paper. These probabilities are asymmetric, meaning that the likelihood of term A appearing given term B is not necessarily the same as the likelihood of term B appearing given term A. For example, a large proportion of papers using LLM terms in this study also mentioned *underscore*, with co-occurrence probabilities ranging from 44% for *foster* to 59% for *delve* (first row in Figure 11), whereas the reverse probabilities were much lower. In contrast, in 2022 these co-occurrence rates were much lower, usually between 1% and 14% across most terms (first row of Figure 10). Similarly, papers using *intricate*, *pivotal*, or *meticulous* in 2024 often mentioned other LLM terms within the same publication, with co-occurrence probabilities between 30% and 45%. In 2022, however, the same terms co-occurred in fewer than 8% of papers. The increased co-occurrence of LLM terms after ChatGPT release suggests their growing usage in academic writing, likely influenced by LLMs.

| | underscore | delve | showcase | unveil | intricate | meticulous | pivotal | heighten | nuance | bolster | foster | interplay |
|---|---|---|---|---|---|---|---|---|---|---|---|---|
| underscore | | 7.46 | 7.01 | 5.44 | 5.99 | 3.67 | 5.06 | 10.72 | 12.15 | 13.98 | 7.70 | 7.13 |
| delve | 1.30 | | 2.21 | 1.65 | 1.87 | 1.45 | 1.01 | 1.55 | 2.53 | 2.26 | 1.75 | 1.21 |
| showcase | 2.21 | 4.02 | | 2.75 | 3.22 | 1.81 | 1.51 | 2.11 | 3.28 | 3.50 | 2.77 | 2.21 |
| unveil | 2.51 | 4.38 | 4.02 | | 4.91 | 2.13 | 4.32 | 2.05 | 1.96 | 2.60 | 2.86 | 4.53 |
| intricate | 3.37 | 6.03 | 5.74 | 5.99 | | 3.51 | 4.69 | 3.65 | 4.02 | 4.44 | 3.18 | 7.10 |
| meticulous | 1.05 | 2.39 | 1.64 | 1.32 | 1.79 | | 1.18 | 1.20 | 1.20 | 1.41 | 1.00 | 11.73 |
| pivotal | 8.21 | 9.41 | 7.74 | 15.17 | 13.52 | 6.65 | | 8.20 | 6.04 | 8.34 | 9.00 | 12.13 |
| heighten | 8.38 | 6.98 | 5.22 | 3.48 | 5.08 | 3.27 | 3.96 | | 11.60 | 13.36 | 7.47 | 5.72 |
| nuance | 7.99 | 9.55 | 6.83 | 2.79 | 4.69 | 2.74 | 2.45 | 9.75 | | 12.15 | 7.74 | 5.22 |
| bolster | 3.56 | 3.30 | 2.82 | 1.44 | 2.01 | 1.25 | 1.31 | 4.35 | 4.71 | | 3.67 | 1.77 |
| foster | 11.96 | 15.58 | 13.61 | 9.63 | 8.78 | 5.38 | 8.62 | 14.83 | 10.13 | 22.35 | | 9.78 |
| interplay | 9.10 | 8.86 | 8.94 | 12.54 | 16.10 | 4.24 | 9.55 | 9.33 | 10.13 | 8.87 | 8.04 | |

Figure 10. Heatmap of co-occurrence probabilities of LLM-associated terms in PMC full texts in 2022 (before ChatGPT's public release). Each cell shows the conditional probability that a paper containing the row term also contains the column term, so probabilities are asymmetric. Darker colours represent higher probabilities.

<table>
<tr><th></th><th>underscor</th><th>delve[s/d/i</th><th>showcase|</th><th>unveil[s/d/</th><th>intricat[e/i</th><th>meticulous[ly]</th><th>pivotal</th><th>heighten[e</th><th>nuance[|d</th><th>bolster[ed,</th><th>foster[|ed,</th><th>interplay[e</th></tr>
<tr><td>underscore[s/d/ing]</td><td></td><td>59.3</td><td>52.0</td><td>51.6</td><td>52.2</td><td>53.7</td><td>49.2</td><td>51.4</td><td>54.5</td><td>57.2</td><td>44.0</td><td>47.8</td></tr>
<tr><td>delve[s/d/ing]</td><td>16.10</td><td></td><td>24.4</td><td>25.5</td><td>23.2</td><td>22.5</td><td>18.6</td><td>18.1</td><td>21.7</td><td>26.0</td><td>17.8</td><td>18.0</td></tr>
<tr><td>showcase[s/(</td><td>13.56</td><td>23.4</td><td></td><td>19.4</td><td>18.5</td><td>20.9</td><td>14.5</td><td>12.5</td><td>15.9</td><td>21.0</td><td>12.8</td><td>11.6</td></tr>
<tr><td>unveil[s/d/ing</td><td>10.89</td><td>19.8</td><td>15.7</td><td></td><td>16.8</td><td>14.1</td><td>14.4</td><td>13.0</td><td>10.7</td><td>15.8</td><td>9.9</td><td>14.0</td></tr>
<tr><td>intricat[e/ies/</td><td>27.31</td><td>44.6</td><td>37.0</td><td>41.7</td><td></td><td>36.7</td><td>33.7</td><td>30.6</td><td>32.4</td><td>36.5</td><td>25.1</td><td>39.3</td></tr>
<tr><td>meticulous[ly</td><td>14.08</td><td>21.7</td><td>21.0</td><td>17.4</td><td>18.4</td><td></td><td>15.2</td><td>14.5</td><td>17.4</td><td>21.0</td><td>11.9</td><td>11.7</td></tr>
<tr><td>pivotal</td><td>34.74</td><td>48.4</td><td>39.1</td><td>48.2</td><td>45.5</td><td>41.0</td><td></td><td>37.5</td><td>33.5</td><td>47.1</td><td>33.9</td><td>37.3</td></tr>
<tr><td>heighten[ed/s</td><td>23.34</td><td>30.3</td><td>21.8</td><td>28.0</td><td>26.6</td><td>25.2</td><td>24.1</td><td></td><td>24.2</td><td>33.2</td><td>23.6</td><td>25.2</td></tr>
<tr><td>nuance[|d/s]</td><td>16.69</td><td>24.4</td><td>18.7</td><td>15.5</td><td>19.0</td><td>20.3</td><td>14.5</td><td>16.3</td><td></td><td>23.5</td><td>19.7</td><td>18.9</td></tr>
<tr><td>bolster[ed/s/i</td><td>7.59</td><td>12.7</td><td>10.7</td><td>9.9</td><td>9.3</td><td>10.6</td><td>8.9</td><td>9.7</td><td>10.2</td><td></td><td>11.3</td><td>7.6</td></tr>
<tr><td>foster[|ed/s/i</td><td>19.44</td><td>29.0</td><td>21.7</td><td>20.7</td><td>21.2</td><td>20.0</td><td>21.2</td><td>23.0</td><td>27.7</td><td>37.6</td><td></td><td>20.5</td></tr>
<tr><td>interplay[ed/s</td><td>21.40</td><td>29.6</td><td>19.9</td><td>29.8</td><td>33.6</td><td>20.0</td><td>23.7</td><td>24.8</td><td>27.7</td><td>25.5</td><td>20.8</td><td></td></tr>
</table>

Figure 11. Heatmap of co-occurrence probabilities of LLM-associated terms in PMC full texts in 2024 (after ChatGPT's public release). Each cell shows the conditional probability that a paper containing the row term also contains the column term, so probabilities are asymmetric. Darker colours represent higher probabilities.

*Correlations between LLM-associated terms (2022 vs. 2024)*

Figures 12 and 13 compare the Pearson correlations between LLM-associated terms in PMC full-text papers. In 2022, the Pearson correlations between most terms were very low, mostly below 0.05. For example, *underscore* had weak correlations with *delve* (r=0.018), *showcase* (r=0.014), and *intricate* (r=0.020), indicating that these terms were rarely used together within the same papers before ChatGPT release. In contrast, there were much stronger correlations across almost all terms. In 2024, *underscore* had the highest overall correlations with other LLM-associated terms, such as *pivotal* (r=0.449), *intricate* (r=0.405), and *nuance* (r=0.338), compared with 2022, where the same correlations were much weaker (0.032, 0.020, and 0.063, respectively). Similarly, *intricate* correlated highly with *delve* (0.335) and *interplay* (0.423), whereas these values were below 0.03 in 2022. These findings indicate that almost two years after the release of ChatGPT academic publications increasingly used multiple LLM-associated terms together.

| | underscore | delve | showcase | unveil | intricate | meticulous | pivotal | heighten | nuance | bolster | foster | interplay |
|---|---|---|---|---|---|---|---|---|---|---|---|---|
| underscore | | 0.018 | 0.014 | 0.017 | 0.02 | 0.004 | 0.032 | 0.059 | 0.063 | 0.047 | 0.03 | 0.04 |
| delve | 0.018 | | 0.016 | 0.019 | 0.029 | 0.012 | 0.017 | 0.018 | 0.042 | 0.031 | 0.018 | 0.02 |
| showcase | 0.014 | 0.016 | | 0.016 | 0.029 | 0.006 | 0.008 | 0.009 | 0.025 | 0.013 | 0.02 | 0.02 |
| unveil | 0.017 | 0.019 | 0.016 | | 0.041 | 0.007 | 0.061 | 0.01 | 0.006 | 0.009 | 0.012 | 0.053 |
| intricate | 0.02 | 0.029 | 0.029 | 0.041 | | 0.013 | 0.045 | 0.018 | 0.025 | 0.015 | 0.011 | 0.084 |
| meticulous | 0.004 | 0.012 | 0.006 | 0.007 | 0.013 | | 0.012 | 0.004 | 0.007 | 0.006 | 0.002 | 0.002 |
| pivotal | 0.032 | 0.017 | 0.008 | 0.061 | 0.045 | 0.012 | | 0.019 | 0.008 | 0.011 | 0.018 | 0.061 |
| heighten | 0.059 | 0.018 | 0.009 | 0.01 | 0.018 | 0.004 | 0.019 | | 0.069 | 0.052 | 0.036 | 0.033 |
| nuance | 0.063 | 0.042 | 0.025 | 0.006 | 0.025 | 0.007 | 0.008 | 0.069 | | 0.054 | 0.049 | 0.037 |
| bolster | 0.047 | 0.031 | 0.013 | 0.009 | 0.015 | 0.006 | 0.011 | 0.052 | 0.054 | | 0.041 | 0.018 |
| foster | 0.03 | 0.018 | 0.02 | 0.012 | 0.011 | 0.002 | 0.018 | 0.036 | 0.049 | 0.041 | | 0.021 |
| interplay | 0.04 | 0.02 | 0.02 | 0.053 | 0.084 | 0.002 | 0.061 | 0.033 | 0.037 | 0.018 | 0.021 | |

Figure 12. Pearson correlation heatmap of LLM-associated terms within PMC full-text papers in 2022 (before ChatGPT's public release), showing associations between term frequencies within the same paper.

| | underscore | delve | showcase | unveil | intricate | meticulous | pivotal | heighten | nuance | bolster | foster | interplay |
|---|---|---|---|---|---|---|---|---|---|---|---|---|
| underscore | | 0.311 | 0.272 | 0.233 | 0.405 | 0.29 | 0.449 | 0.295 | 0.338 | 0.23 | 0.224 | 0.315 |
| delve | 0.311 | | 0.218 | 0.211 | 0.335 | 0.211 | 0.295 | 0.182 | 0.231 | 0.168 | 0.162 | 0.2 |
| showcase | 0.272 | 0.218 | | 0.144 | 0.256 | 0.222 | 0.21 | 0.098 | 0.159 | 0.128 | 0.085 | 0.095 |
| unveil | 0.233 | 0.211 | 0.144 | | 0.316 | 0.166 | 0.302 | 0.172 | 0.098 | 0.102 | 0.068 | 0.192 |
| intricate | 0.405 | 0.335 | 0.256 | 0.316 | | 0.298 | 0.491 | 0.255 | 0.274 | 0.186 | 0.147 | 0.423 |
| meticulous | 0.29 | 0.211 | 0.222 | 0.166 | 0.298 | | 0.251 | 0.136 | 0.197 | 0.146 | 0.084 | 0.118 |
| pivotal | 0.449 | 0.295 | 0.21 | 0.302 | 0.491 | 0.251 | | 0.285 | 0.195 | 0.243 | 0.192 | 0.309 |
| heighten | 0.295 | 0.182 | 0.098 | 0.172 | 0.255 | 0.136 | 0.285 | | 0.14 | 0.164 | 0.146 | 0.211 |
| nuance | 0.338 | 0.231 | 0.159 | 0.098 | 0.274 | 0.197 | 0.195 | 0.14 | | 0.135 | 0.192 | 0.214 |
| bolster | 0.23 | 0.168 | 0.128 | 0.102 | 0.186 | 0.146 | 0.243 | 0.164 | 0.135 | | 0.195 | 0.119 |
| foster | 0.224 | 0.162 | 0.085 | 0.068 | 0.147 | 0.084 | 0.192 | 0.146 | 0.192 | 0.195 | | 0.13 |
| interplay | 0.315 | 0.2 | 0.095 | 0.192 | 0.423 | 0.118 | 0.309 | 0.211 | 0.214 | 0.119 | 0.13 | |

Figure 13. Pearson correlation heatmap of LLM-associated terms within PMC full-text papers in 2024 (after ChatGPT's public release), showing associations between term frequencies within the same paper.

## Discussion

This study is the first large-scale analysis to assess both the frequency and co-occurrence of LLM-associated terms in the full texts of academic publications following the release of ChatGPT. Unlike previous studies that primarily counted single mentions of terms in academic publications (e.g., in titles or abstracts), we examined how frequently LLM-associated terms were used and how often they co-occurred within the full text of the same paper. This approach provides a more detailed understanding of how LLMs have influenced academic writing.

Our analysis showed that *underscore* had the highest frequency of use, the strongest co-occurrence with other LLM-associated terms, and the highest correlations within PMC full texts after ChatGPT's release. One reason for the prominence of *underscore* is that it might be more acceptable in academic writing. LLMs frequently suggest "underscore" when generating abstracts, editing, or proofreading, and many authors may keep it unchanged because it sounds appropriate within academic texts (e.g., "These findings underscore the importance…"). In contrast, terms like delve or showcase seem to be less academic and more likely to be replaced or rephrased by authors, or avoided by LLMs detecting the academic context of the user prompt.

Some LLM-associated terms may be more useful for titles, abstracts, or full-texts, depending on how authors and LLMs use them. For example, *unveil* appears to be highly title-friendly, with the largest relative increase in Scopus titles between 2022 and 2024 (from 0.04% to 0.26%). In contrast, *underscore* is used widely across abstracts, full texts, and multiple sections of papers, making it one of the most commonly used LLM-suggested terms. *Delve* appears to be more abstract-friendly than full-text-friendly, as shown by its higher growth in abstract databases but relatively limited repeated use within main texts (Figure 2 and 6). These findings indicate that LLMs may influence different parts of academic papers differently, with some terms being more likely to appear in titles (see also Comas-Forgas, Koulouris, & Kouis, 2025) or abstracts while others are used more broadly throughout the full text.

**Are LLM-associated terms increasing faster than traditional research terms?**

A follow-up analysis conducted on 28 January 2025 confirmed that the use of LLM-associated terms in titles, abstracts, and keywords of Scopus papers continues to rise sharply. In contrast, the frequency of a set of traditional academic terms with similar meanings (used here as control terms) has remained relatively stable over the same period (Table 2). To assess whether the observed increases were specific to LLM-associated terms, we compared them with more traditional alternatives that have similar meanings and are well established in academic writing. These comparison terms were chosen pragmatically, focusing on widely used semantically similar terms with stable usage across disciplines (e.g., investigate as a traditional counterpart to delve). Other reasonable alternatives, such as assess, examine, explore, or probe, were not tested in this context, as the aim was to keep the comparison practical and easy to interpret for discussion, rather than covering all possible alternatives. For noun phrases, we used academic terms with close semantic overlap (e.g., interaction as a conventional counterpart to interplay), although future research could investigate a wider range of alternative terms.

For example, between 2022 and 2024, the LLM-associated term *delve* increased by 1,360%, compared with only an 8.9% increase for the traditional term *investigate*. Similarly, *underscore* rose by 1,062%, while the comparable term *highlight* increased by just 85%. Other LLM-associated terms, such as *intricate* (+727%) and *meticulous* (+611%), also grew far faster than their traditional counterparts *complex* (+18%) and *precise* (+68%). This supports, but does not prove, the hypothesis that LLMs are the cause of the differences rather than changes in what scientists have written about, or lengthening abstracts (which would make all terms more common).

Table 2. Percentage increase in the use of LLM-associated terms versus comparable traditional academic terms in Scopus papers (2022–2024). Note that the high percentage increases for some terms are only possible because of the low baseline values.

| **Common term** | 2022 (%) | 2024 (%) | **% Increase** | **LLM term** | 2022(%) | 2024 (%) | **% Increase (2022-2024)** |
|---|---|---|---|---|---|---|---|
| investigate | 17.83 | 19.41 | **8.90%** | delve | 0.07 | 1.05 | **1360%** |
| highlight | 5.08 | 9.43 | **85.68%** | underscore | 0.25 | 2.87 | **1062%** |
| demonstrate | 14.07 | 20.35 | **44.72%** | showcase | 0.20 | 0.99 | **395%** |
| reveal | 12.03 | 16.02 | **33.25%** | unveil | 0.26 | 0.88 | **235%** |
| complex | 10.02 | 11.78 | **17.63%** | intricate | 0.14 | 1.20 | **727%** |
| precise | 1.74 | 2.93 | **67.94%** | meticulous | 0.06 | 0.45 | **611%** |
| critical | 6.45 | 7.99 | **23.92%** | pivotal | 0.40 | 1.62 | **308%** |
| enhance | 9.56 | 18.76 | **96.25%** | heighten | 0.15 | 0.57 | **273%** |
| detail | 4.16 | 4.37 | **4.94%** | nuanced | 0.20 | 0.61 | **210%** |
| strengthen | 1.48 | 1.65 | **11.42%** | bolster | 0.06 | 0.27 | **361%** |
| promote | 5.87 | 6.99 | **19.17%** | foster | 0.50 | 1.40 | **177%** |
| interaction | 8.25 | 9.06 | **9.89%** | interplay | 0.45 | 0.99 | **119%** |

**Are LLM-associated terms more common in retracted articles?**

Figure 14 compares the percentage of retracted and published research articles that included any of the 12 LLM-associated terms in Scopus titles or abstracts from 2015 to 2024. The results show a clear rise in the use of these terms in retracted articles after the release of ChatGPT in 2022. In 2024, about 15.8% (262 of 1,657) of retracted papers used at least one LLM term, nearly double the 8.1% (245,741 of 3,035,375) found in published papers. However, this difference should be interpreted as a hypothesis-generating observation rather than evidence of a causal relationship, because retractions are uncommon and may reflect a wide range of quality factors unrelated to writing practices. Moreover, it does not detract from the entirely legitimate use of LLM translations in academic writing.

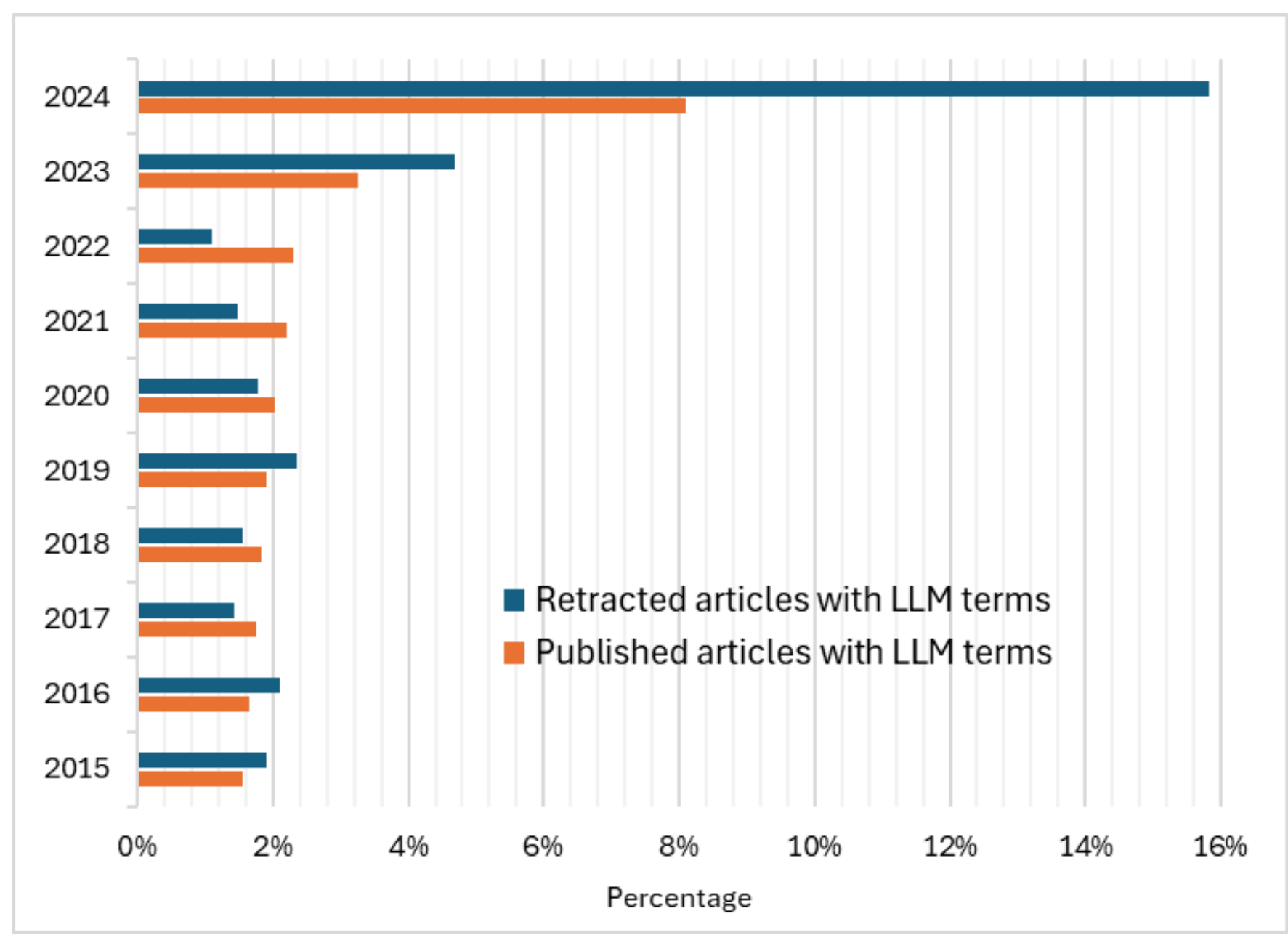

Figure 14. Percentage of retracted and published research articles in Scopus (2015–2024) that included any of 12 LLM-associated terms in the title or abstract.

**Limitations**

The results are limited by the set of 12 LLM-associated terms used and the six databases analysed, and we may have overlooked other terms commonly suggested by LLMs, including those used by models other than ChatGPT. The findings may also change in the future as LLMs evolve, for example, if DeepSeek or other emerging models largely replace existing systems. Moreover, some of the observed terms may reflect existing word choices in earlier LLM training data and may change over time as models adapt to user feedback. Our analysis is based on English-language terms and mostly on metadata (e.g., titles and abstracts), which may introduce biases. For instance, non-English articles indexed with translated English abstracts could contain LLM-associated terms even if the original manuscript does not. Moreover, non-native English-speaking authors may be more likely to use LLMs for proofreading, editing, and translation to improve clarity, which could influence LLM-associated term counts and frequencies.

Another limitation is that some of the studied terms may naturally be used more frequently in specific disciplines, independent of LLM usage. For example, the term *foster* or *fostering* may commonly appear in fields like education, social work, child development or animal care, making these counts less reliable as direct indicators of LLM influence. Similarly, *heighten* may be more common in disciplines like psychology and neuroscience (e.g., “heightened anxiety” or “heightened stress”). In contrast, terms like *delve* and *underscore* are less domain-specific and therefore more likely to indicate stylistic patterns influenced by LLMs.

The PMC full-text dataset is available only for open-access biomedical and life-science research, and not all articles are included in the analysis. As a result, the findings mainly reflect LLM-term usage in open-access publications and may not fully represent non-open-access journals including many leading medical journals such as The New England Journal of Medicine, The Lancet, and JAMA or other disciplines.

Finally, this study analysed changes in the frequency and co-occurrence of selected LLM-associated words in academic publications but did not assess potential semantic or stylistic changes in sentence structure which may also be influenced by LLMs.

## Conclusions

**In answer to RQ1**, the use of LLM-associated terms has increased sharply across scholarly databases after ChatGPT's public release in late 2022. *Delve* (+1,360%) and *underscore* (+1,062%) grew much faster than traditional terms such as *investigate* (+8.9%) and *highlight* (+85%) between 2022 and 2024. By 2024, *underscore* appeared in about 20% of PMC full-text papers and 11% of Dimensions papers, while *pivotal* reached 15% in PMC. There are also clear disciplinary differences. The growth of LLM terms was highest in STEM fields, often exceeding 3,000%. In contrast, there were smaller increases in the social sciences and arts and humanities, mostly below 500%. Some terms, such as *delve*, were more common in arts, social sciences, and business, while *underscore* dominates in STEM disciplines.

**In answer to RQ2**, the PMC full-text analysis showed that the LLM-associated terms studied are not only appearing more often but are also being used more frequently within publications. For example, papers using *underscore* six or more times increased by over 10,000% from 2022 to 2025 (see Figure 9 for the underlying percentages). Both *intricate* and *meticulous* also had strong growth in repeated usage, with increases of 5,400% and 2,800%, respectively. By contrast, although *delve* had the highest growth in abstract databases such as Scopus and Web of Science (Figure 2), the full-text analysis showed that it is less likely to appear very frequently in the main body of papers. On average, it was used up to five times per paper, suggesting that *delve* is a more LLM abstract-friendly term (e.g., "This study delves into…") rather than one that is repeatedly useful within the main text (Figure 6).

**In answer to RQ3,** the co-occurrence analysis of PMC full texts shows that LLM-associated terms are increasingly used together within the same papers after the release of ChatGPT. In 2024, 59.3% of papers mentioning *delve* (37,534 in total) also included *underscore* (22,241 papers), compared with just 1.3% in 2022. Similarly, *intricate*, *pivotal*, and *meticulous* co-occurred with other LLM-associated terms in 30%–45% of PMC papers by 2024, while in 2022 their co-occurrence rates were typically below 8%. This suggests a growing co-occurrence of LLM-terms in academic writing and editing practices.

**In answer to RQ4**, the correlation analysis of PMC full texts shows that LLM-associated terms became much more closely related in 2024 compared to 2022. In 2024, correlations between the studied LLM-associated terms were much stronger (e.g., *underscore* with *delve*: $r = 0.311$; *underscore* with *pivotal*: $r = 0.449$) compared with 2022 ($r = 0.018$ and $r = 0.032$, respectively). *Underscore* had the highest overall correlations with other LLM-associated terms (ranging from 0.230 to 0.449) in 2024, compared with 2022 (ranging only 0.004 to 0.063). This shows that papers using one LLM-associated term are now much more likely to include other terms within the same publication after ChatGPT's release.

Although this study provides new evidence that LLMs like ChatGPT are associated with changes in academic writing through the analysis of updated data, a broader range of terms, and multiple scholarly databases, further research is needed to understand how LLMs are shaping academic publishing across specific subjects, considering their relatively recent introduction. Different LLMs (e.g., ChatGPT, Gemini, and DeepSeek) may also use other terms when generating or editing academic texts. Hence, future research could investigate differences between LLMs in their influence on academic writing. In particular, the study cannot determine whether these patterns reflect LLMs producing text directly for editing, or from wider changes in editorial or publishing practices linked to LLM use. Hence, the findings should be interpreted as evidence of changing language patterns rather than as direct proof of how LLMs are used in producing academic texts. Moreover, the analysis focuses on English-language terms and patterns in other languages remain unclear.

Finally, despite concerns about the misuse of AI suggested by the retraction association above, and prior studies (Cabanac et al., 2021), the increasing prevalence of AI-related language in full texts as well as titles and abstracts may be a welcome development overall. It suggests that LLMs are reducing the language barrier to academic publishing for non-English speakers and hence partly addressing the current unacceptable level of global inequality in science publishing. Nevertheless, since most of the world does not speak English, this suggests that the volume of publishable academic material produced annually is rising and is likely to continue to increase rapidly; this must put considerable pressure on the current publishing system. For example, it might lead to higher rejection rates, more and/or larger journals, or alternative publishing strategies.

**Funding:** No funding was provided for this study.

**Data availability:** The shared dataset includes the frequency of LLM-associated terms within PMC papers from 2021 to July 2025, along with the data and statistics used in the analyses reported in this study. The dataset is available via https://doi.org/10.6084/m9.figshare.30102913

**Declarations**

**Competing interests:** The authors are members of the Distinguished Reviewers Board of Scientometrics.

Appendix A. Percentage of academic publications containing six additional LLM-associated terms across six databases from 2015 to 2024 (complementary to Figure 1).

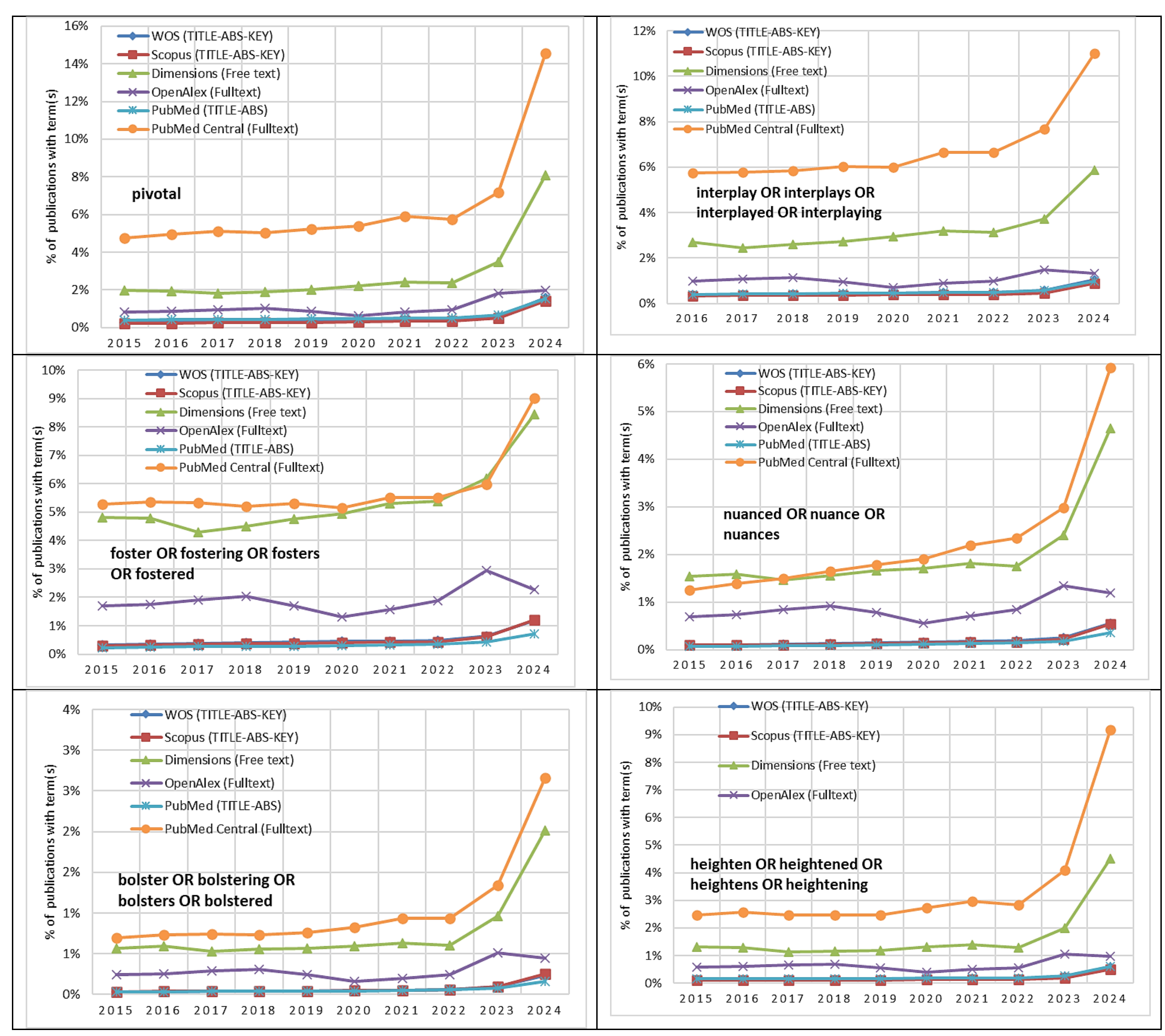